\documentclass{article}
\usepackage{smc}
\usepackage{times}
\usepackage{ifpdf}
\usepackage[english]{babel}
\usepackage{cite}

\usepackage{amsmath}
\DeclareMathOperator*{\argmax}{argmax}

\def\papertitle{A DATA-DRIVEN METHODOLOGY FOR CONSIDERING FEASIBILITY AND PAIRWISE LIKELIHOOD IN DEEP LEARNING BASED GUITAR TABLATURE TRANSCRIPTION SYSTEMS}
\def\firstauthor{Frank Cwitkowitz}
\def\secondauthor{Jonathan Driedger}
\def\thirdauthor{Zhiyao Duan}

\newif\ifpdf
\ifx\pdfoutput\relax
\else
   \ifcase\pdfoutput
      \pdffalse
   \else
      \pdftrue
\fi

\ifpdf 
  \usepackage[pdftex,
    pdftitle={\papertitle},
    pdfauthor={\firstauthor, \secondauthor, \thirdauthor},
    bookmarksnumbered, 
    pdfstartview=XYZ 
   ]{hyperref}

  \usepackage[pdftex]{graphicx}
  \graphicspath{{./figures/}}
  \DeclareGraphicsExtensions{.pdf,.jpeg,.png}

  \usepackage[figure,table]{hypcap}

\else 
  \usepackage[dvips,
    bookmarksnumbered, 
    pdfstartview=XYZ 
  ]{hyperref}  

  \usepackage[dvips]{epsfig,graphicx}
  \graphicspath{{./figures/}}
  \DeclareGraphicsExtensions{.eps}

  \usepackage[figure,table]{hypcap}
\fi

\hypersetup{
    colorlinks,%
    citecolor=black,%
    filecolor=black,%
    linkcolor=black,%
    urlcolor=black
}

\title{\papertitle}

 \threeauthors
   {\firstauthor $^\dagger$\thanks{$\dagger$ Main work 
   completed as a research intern at Chordify.}} {University of Rochester \\ %
     {\tt \href{mailto:fcwitkow@ur.rochester.edu}{fcwitkow@ur.rochester.edu}}}
   {\secondauthor} {Chordify \\ %
     {\tt \href{mailto:jonathan@chordify.net}{jonathan@chordify.net}}}
   {\thirdauthor} {University of Rochester \\ %
     {\tt \href{mailto:zhiyao.duan@rochester.edu}{zhiyao.duan@rochester.edu}}}

\begin{document}
\capstartfalse
\maketitle
\capstarttrue
\begin{abstract}
Guitar tablature transcription is an important but understudied problem within the field of music information retrieval. Traditional signal processing approaches offer only limited performance on the task, and there is little acoustic data with transcription labels for training machine learning models. However, guitar transcription labels alone are more widely available in the form of tablature, which is commonly shared among guitarists online. In this work, a collection of symbolic tablature is leveraged to estimate the pairwise likelihood of notes on the guitar. The output layer of a baseline tablature transcription model is reformulated, such that an inhibition loss can be incorporated to discourage the co-activation of unlikely note pairs. This naturally enforces playability constraints for guitar, and yields tablature which is more consistent with the symbolic data used to estimate pairwise likelihoods. With this methodology, we show that symbolic tablature can be used to shape the distribution of a tablature transcription model's predictions, even when little acoustic data is available.
\end{abstract}

\section{Introduction}\label{sec:introduction}
Automatic Music Transcription (AMT) is a well-known task within the Music Information Retrieval (MIR) community dealing with the estimation of note content within a music signal \cite{benetos2018automatic}. Guitar tablature transcription refers to the specific problem of estimating all of the notes within a solo guitar recording and identifying the strings that were used to play them. The task represents the determination of not only what was played, but also how it was played on the instrument. This information is necessary to realize tablature, a type of prescriptive notation for stringed instruments where fret numbers are superimposed atop lines representing each string. The fret numbers correspond to the notes that are to be played for a specific piece.

The guitar is a very popular musical instrument with users spanning all skill levels. The value of knowing how a piece was played on guitar is immeasurable for the vast community of guitarists learning to play the instrument. Guitar has a relatively low barrier to entry w.r.t. music theory knowledge, and many guitarists use guitar tablature instead of standard staff notation. Even more experienced players often prefer tablature for storing and communicating guitar-specific music ideas due to its intuitiveness and simplicity. Tablature is also widely shared across the internet through primarily user-curated websites such as \textit{Ulimate-Guitar}\footnote{\url{https://www.ultimate-guitar.com/}}.

Despite the popularity of guitar, the instrument has received considerably less attention when it comes to music transcription. The main obstacles stem from a lack of audio recordings with transcription labels, referred to here as acoustic data, capturing the exceeding variability of the instrument. The guitar has many expressive dimensions such as the plucking style, plucking location, the use of embellishments\footnote{Examples include bends, slides, hammer-ons/offs, vibrato, etc.}, etc. These and many other factors can affect the audio. Standard guitars do not have a digital interface, and expensive manual processing is required to obtain qualitative note labels. Without large datasets capturing the breadth of the intrument, it is very difficult to train reliable models and avoid over-fitting.

Since a standard guitar has six independent strings, it is a polyphonic instrument. This means that guitar transcription carries all of the intrinsic challenges of polyphonic note transcription. One further challenge is the estimation of the string on which each note was played. In standard tuning, the pitch ranges of adjacent strings have a significant overlap, with their lowest pitches being only 4-5 semitones apart. The timbral cues of different strings are subtle, and it is difficult to learn these without a lot of data.

There have been several attempts to realize systems which can transcribe solo guitar audio into tablature. Often, the approaches consist of two-stage systems, which first estimate pitch salience and then map pitch estimates to the guitar. Many works employ a basic signal processing pipeline, \textit{e.g.} \cite{fiss2011automatic, alcabasa2012automatic, kehling2014automatic}, whereby some form of spectral analysis or peak-picking is carried out. Some approaches attempt to estimate the string of detected pitches based off of inharmonicity measurements \cite{barbancho2012inharmonicity, michelson2018automatic}, which vary across strings. Others employ graph-search algorithms to find the optimal path through a list of possible fingerings for the observed pitches \cite{burlet2013robotaba, yazawa2013audio, yazawa2014automatic}. Often, to determine if a fingering or a transition is optimal or even feasible, these approaches rely on rule-based procedures with hand-crafted weightings.

More recently, machine learning has become a popular strategy for guitar tablature transcription. Several works train Hidden Markov Models (HMMs) to model the transition between fingerings and chords \cite{barbancho2011automatic, hori2013input}. Other works perform classification, \textit{e.g.} \cite{kehling2014automatic}, which uses a collection of hand-crafted features to estimate, among other parameters, the string associated with each note, or \cite{burlet2017isolated}, where a Deep Belief Network (DBN) is utilized to produce pitch and polyphony estimates. Bayesian classification has also been proposed \cite{michelson2018automatic, hjerrild2019estimation, hjerrild2019physical} to estimate the string and fret of notes.

Convolutional Neural Network (CNN) based approaches have been proposed \cite{humphrey2014music, wiggins2019guitar} to perform the task of pitch and string estimation jointly. These models benefit from being able to learn features directly from acoustic data, and tend to generalize much more effectively to real-world data. Our main contributions stem from a simple observation: the output layer formulation of these CNNs is prone to falsely producing tablature with duplicated pitches. This is because the output layer is formulated as six independent classification problems, \textit{i.e.}, one per string. While the output neurons of each softmax group implicitly share information from previous layers of the network, they do not explicitly communicate when determining which class to choose. In contrast, the fretting of one string is highly correlated to the fretting of other strings. This relationship is based on what pitch intervals are likely to be played at different locations as well as bio-mechanical feasibility.

In order to incorporate the knowledge of likely fingerings and feasibility, we propose a new output layer formulation for guitar tablature transcription models\footnote{All code is available at \url{https://github.com/cwitkowitz/guitar-transcription-with-inhibition}.}. In the new formulation, a novel inhibition objective is applied during training to discourage the concurrent activation of unlikely or infeasible fingerings. The inhibition weights are derived from the likelihood of co-occurrence for each pair of notes on the guitar, which is estimated using DadaGP \cite{sarmento2021dadagp}, a large dataset of guitar tablature. The proposed output layer formulation essentially learns a language model without requiring acoustic data. We directly compare the new formulation to the previous formulation, and show that its predictions more closely match the distribution of DadaGP.

\section{Proposed Method}
\label{sec:proposed_method}
Given a collection of symbolic tablature data, the pairwise likelihood of each string and fret (S/F) combination can be estimated. The new output layer formulation is amenable to training with the pairwise likelihoods through a pairwise inhibition loss. As a result, the predictions of the new output layer more closely match the distribution of the tablature within the collection. Assuming a preponderance of the tablature is playable, the generated tablature will naturally be more feasible to play. In the following sections, we introduce the baseline CNN architecture used in this work, and discuss the new output layer formulation in more detail, the process of estimating the pairwise likelihood of each S/F combination, and the proposed inhibition loss.

\subsection{Baseline Model}
\label{sec:baseline_model}
We employ TabCNN \cite{wiggins2019guitar} as our baseline model for guitar tablature transcription, leaving most of the original design choices largely unchanged. TabCNN is a simple CNN which processes Constant-Q Transform (CQT) frames and produces sets of fret class predictions. It was designed to be compatible with real-time processing, so it is relatively lightweight and operates on multiple CQT frames in order to make one set of predictions. The model comprises three 2D convolutional layers with ReLU activations, followed by a max pooling layer, a fully-connected layer with ReLU activation, and finally the output layer discussed in the following section. During training, dropout is applied after the max pooling layer and directly before the output layer. The network is trained using AdaDelta optimizer with an initial learning rate of $1.0$. We implemented the model from scratch in PyTorch, using all of the same hyperparameters as in the original paper for feature extraction and the model architecture. We also insert a uni-directional long short-term memory (LSTM) \cite{hochreiter1997long} layer before the output layer. This is a simple modification which results in a relatively significant improvement (see Sec. \ref{sec:results_and_discussion}), without disrupting the real-time processing capacity of TabCNN.

\subsection{Output Layer Formulation}
\label{sec:output_layer_formulation}
The output layer of TabCNN is a fully-connected layer with one softmax activation for each string. The output neurons represent all combinations of string $s \in \{1, ..., 6\}$ and fret class $f \in \{-1, 0, 1, ..., F\}$, where $f = -1$ represents a class for silence, $f = 0$ represents the open string, and $F$ is the total number of frets supported. In the standard model, the softmax operations are applied independently to the neurons associated with each string. This results in probability activations $z_{s, f, n}$ for each frame $n$ where $z_{s, f, n} > 0~\forall s, f, n$ and $\sum_{f = -1}^{F} z_{s, f, n} = 1,~\forall s, n$. The loss for a track is then computed by summing the categorical cross entropy for each string group and averaging across the $N$ total frames. This can be expressed as
\begin{equation}
\label{eq:categorical_cross_enropy}
L_{CCE} = -\frac{1}{N} \sum_{n = 1}^{N} \sum_{s = 1}^{6} \log{(z_{s, f', n})},
\end{equation}
where $z_{s, f', n}$ is the activation corresponding to the ground truth fret $f'$ for string $s$ during frame $n$. Inference then consists of choosing the frets with the highest activation, resulting in six predictions for each frame. Using a binary representation, the predictions can be written as
\begin{equation}
\label{eq:tablature_inference}
y_{s, f, n} = \mathcal{I}(\argmax_{f} \{z_{s, f, n}\} = f),
\end{equation}
where $\mathcal{I(\cdot)}$ is the indicator function. We refer to this formulation as the 6D Softmax formulation. One advantage of this formulation is that the model is not capable of generating invalid predictions, \textit{i.e.}, only one note can be chosen at maximum for each string. While this property is highly desirable, the six softmax activations act independent of one another, causing the model to treat transcription as six somewhat independent classification tasks. In practice, the ground truth S/F combinations making up a fingering arrangement are highly correlated at all times. This is due to both the physical limitations of what a human hand can play as well as musical motivations that make it unlikely to play certain pitches at the same time.

In contrast, we propose to formulate the output layer as representing a binary classification problem for each fret of each string. In this way, the likelihood of each S/F combination being active is independently computed using a sigmoid activation. As such, we refer to the new formulation as the Logistic formulation. This formulation allows us to expand the loss function to include an inhibition objective to discourage the co-activation of certain pairs of S/F combinations. This additional objective would otherwise conflict with the 6D Softmax formulation, due to normalization of the activations in the softmax function.

The new loss is computed by summing the binary cross entropy across each fret class for each string and again averaging across all frames. For convenience, let us now fold string and fret into a single variable $c \in \{1, ..., C\}$ representing each distinct S/F combination, where $C = 6 \times (F + 2)$. The loss can then be expressed as
\begin{align}
\label{eq:binary_cross_enropy}
L_{BCE} = -\frac{1}{N} \sum_{n = 1}^{N} \sum_{c = 1}^{C} t_{c, n} &\log{(z_{c, n})} \text{ } + \nonumber\\
(1 - t_{c, n}) &\log{(1 - z_{c, n})},
\end{align}
where $t_{c, n}$ is a binary number indicating whether there is a positive class label in the ground-truth at the corresponding S/F combination $c$ at frame $n$. This new output layer formulation is similar to that of standard piano transcription models \cite{hawthorne2017onsets}. However, we still use Equation (\ref{eq:tablature_inference}) to obtain the final predictions, rather than considering each activation above a certain threshold a positive prediction. This is because, ultimately, we can still only choose one fret class per string. Note that the inhibition objective is applied during training and is unaffected by inference.

\subsection{Estimating Pairwise Likelihood}
\label{sec:estimating_pairwise_likelihood}
Certain S/F combinations on the guitar have a very low chance of being played at the same time. These can include S/F combinations located on the same string, S/F combinations with the same pitch, S/F combinations which are far apart, or simply musically uncommon S/F combinations. In order to incorporate these considerations into the Logistic formulation, we estimate the likelihood of co-occurrence for all S/F combination pairs, to inform a novel training objective for pairwise S/F combination inhibition.

The pairwise likelihood of two S/F combinations $c_i$ and $c_j$ can be estimated using an arbitrary collection of symbolic tablature data. Here we define symbolic tablature data as one-hot encoded annotations for each string of the guitar at the frame-level. Given the symbolic tablature for a single track, we compute the intersection over union ($IoU$) of frame-level occurrences for all pairs of S/F combinations that co-occur in at least one frame. Mathematically, the intersection of a pair is defined as the number of frames where both combinations occur concurrently:
\begin{equation}
\label{eq:intersection}
inter(i, j) = \sum_{n = 1}^{N} t_{c_i, n} \land t_{c_j, n}.
\end{equation}
The union is defined as the number of frames where either of the combinations occur.
\begin{equation}
\label{eq:union}
union(i, j) = \sum_{n = 1}^{N} t_{c_i, n} \lor t_{c_j, n}.
\end{equation}
Let $\mathcal{T'}(i,j)$ be the set of tracks where $c_i$ and $c_j$, independently, each occur in at least one frame. Then, the $IoU$ of the pair is averaged across these valid tracks:
\begin{equation}
\label{eq:intersection_over_union}
    IoU(i, j) = \frac{1}{|\mathcal{T'}(i,j)|} \sum_{t \in \mathcal{T'}(i,j)} \frac{inter(i, j)_t}{union(i, j)_t},
\end{equation}
where $|\mathcal{T'}(i,j)|$ is the cardinality of $\mathcal{T'}(i,j)$. Note that this is only valid for pairs where $|\mathcal{T'}(i,j)| > 0$. All pairs where $|\mathcal{T'}(i,j)| = 0$ receive $IoU(i, j) = 0$.

\begin{figure*}[t]
\includegraphics[width=0.5\linewidth]{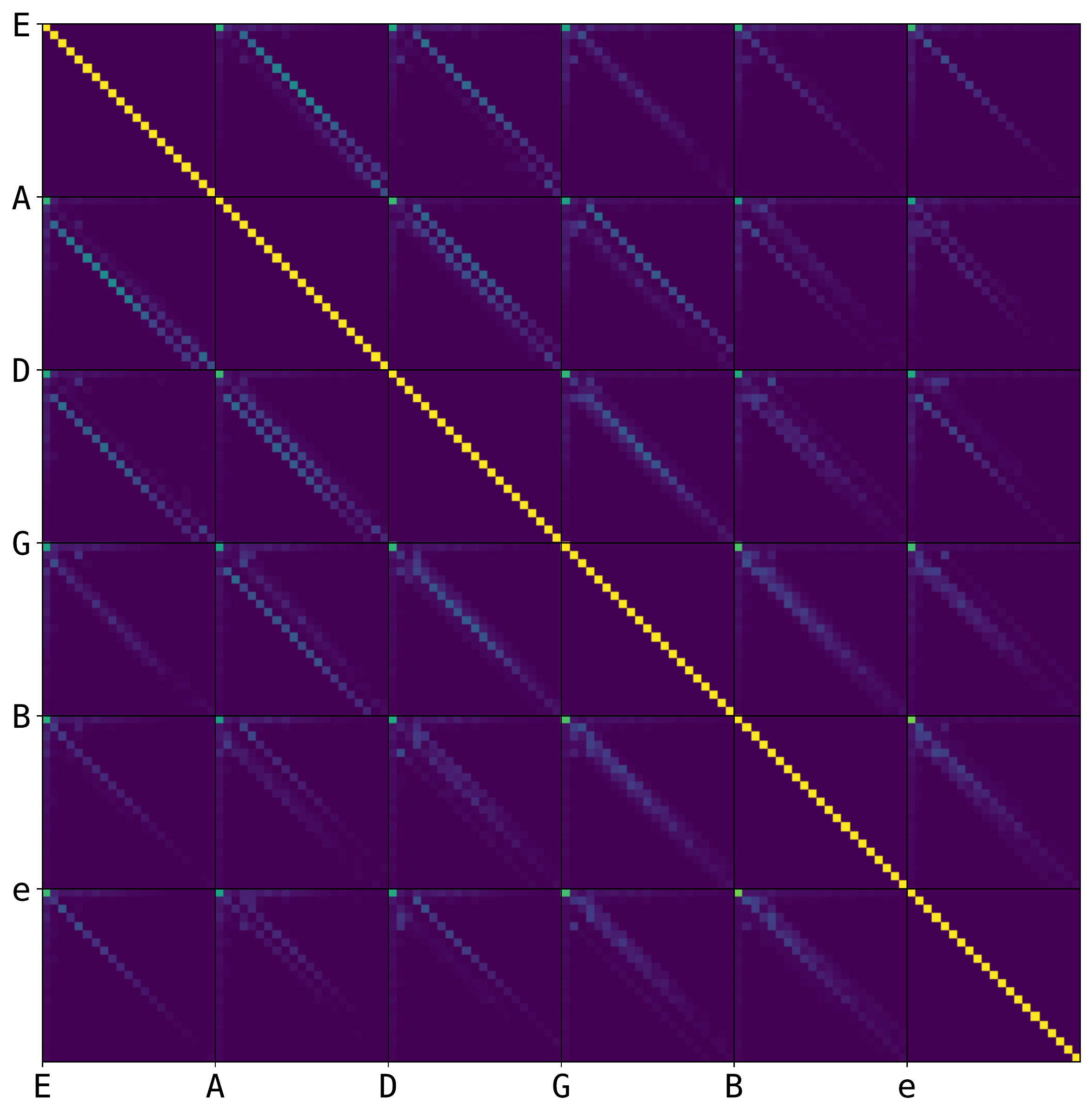}
\includegraphics[width=0.5\linewidth]{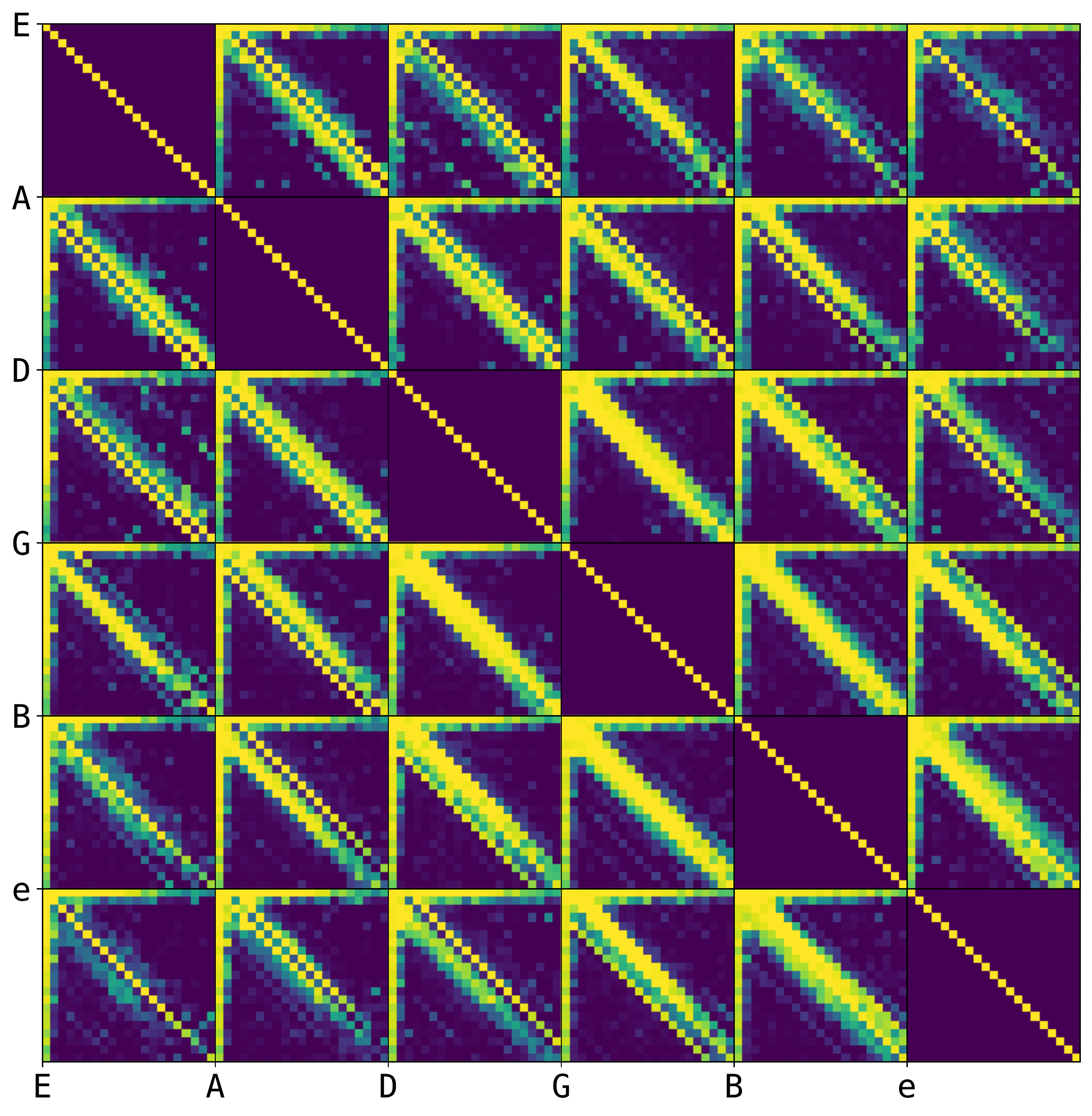}
\caption{Visualization of estimated pairwise likelihood for all S/F combinations, ordered row- and column-wise by string $s$ and fret $f$, computed using DadaGP \cite{sarmento2021dadagp} with $b = 1$ (left) and $b = 2^7$ (right). Grid lines are overlayed atop the string boundaries. The likelihoods range from 0 (dark purple) to 1 (bright yellow), where 1 represents pairs which always co-occur (\textit{e.g.,} identical pairs on diagonal) and 0 represents pairs which never co-occur (\textit{e.g.,} non-identical same-string pairs).}
\label{fig:pairwise_likelihoods_dadagp}
\end{figure*}

The final pairwise likelihoods are stored in a symmetric matrix, ordered by string and fret on both axes. As a result of Equation (\ref{eq:intersection_over_union}), the likelihood of a pair co-occurring with itself is always $1$, and the likelihood of pairs within the same string co-occurring is always $0$. This is convenient for the next step, as self-occurrence will never be inhibited, whereas same-string-occurrence will be maximally inhibited. The pairwise likelihood for the rest of the S/F combinations with differing strings will fall somewhere in between $0$ and $1$ (inclusive). See Fig. \ref{fig:pairwise_likelihoods_dadagp} for an example of a pairwise likelihood matrix.

\subsection{Inhibition Loss}
\label{sec:inhibition_loss}
In order to apply the estimated pairwise likelihoods to the problem of guitar tablature transcription, we introduce a new loss term for inhibiting the co-activation of unlikely pairs. We refer to this as the inhibition loss $L_{inh}$. The inhibition loss requires that each pair of S/F combinations receive an inhibition weight $w(c_i, c_j)$ between $0$ and $1$, indicating how much to penalize the model for producing high activations for the combinations in the pair in a single frame. The inhibition weights here are chosen to be the complement of the pairwise likelihood ($IoU$) estimated in the previous step:
\begin{equation}
\label{eq:inhibition_weight}
w(c_i, c_j) = (1 - IoU(i, j))^{b},
\end{equation}
where $b$ is a parameter which boosts the effective pairwise likelihood by pushing the mass of the inhibition weights closer to $0$. Many S/F combinations which tend to co-occur do not necessarily have a high likelihood of occurring together, relative to the amount of times they occur separately. This can make the estimated pairwise likelihood for many combinations small, leading to a high inhibition weight. Since we do not wish to discourage the activations of pairs which commonly co-occur, it can be helpful to boost the pairwise likelihood in this way. Empirically, we found that $b = 2^7$ produced weights with nice contrast between common pairs and uncommon pairs. Boosting can also be thought of as computing the joint probability of not observing $c_i$ and $c_j$ across $b$ total frames, $\approx 3$ seconds here. Given the inhibition weights, the inhibition loss for a sequence of $N$ frame is computed as
\begin{equation}
\label{eq:inhibition_loss}
L_{inh} = \frac{1}{2N} \sum_{n = 1}^{N} \sum_{i = 1}^{C} \sum_{j = 1}^{C} z_{c_i, n} z_{c_j, n} w(c_i, c_j).
\end{equation}
Here, the product for every combination of activations produced by the model is taken and scaled by the appropriate inhibition weight, and the result is subsequently summed over all combinations. Since both permutations of each S/F combination are included in the summation, we divide by two to remove redundancy. The total loss then becomes
\begin{equation}
\label{eq:total_loss}
L_{total} = L_{BCE} + \lambda L_{inh},
\end{equation}
where $\lambda$ is a scaling term for balancing the two terms.

\section{Experimental Setup}
\label{sec:experimental_setup}
In order to evaluate the efficacy of the proposed output layer formulation, we use it within TabCNN \cite{wiggins2019guitar}. We compare it to the 6D Softmax formulation, and experiment with several variations to study the effect of inhibition.

\subsection{Datasets}
\label{sec:datasets}
We use two datasets for our experiments. GuitarSet \cite{quingyang2019guitarset} is used to train, validate, and test our models, and {DadaGP} \cite{sarmento2021dadagp} is used to compute weights for the inhibition loss applied during training and employed as an evaluation metric. We briefly introduce the datasets in the following sections.

\subsubsection{GuitarSet}
\label{sec:guitarset}
GuitarSet \cite{quingyang2019guitarset} is a guitar transcription dataset comprising roughly three hours of acoustic guitar audio. It contains various types of annotations, including pitch and note annotations with string labels. The labels provided were obtained by employing monophonic pitch tracking on the independent audio of each string, recorded with a hexaphonic pickup mounted to the guitar. The dataset features six guitarists playing two unique interpretations over 30 different chord progressions from various keys, resulting in 360 distinct tracks. Each track is approximately 1312.7 frames or 30.5 seconds on average. GuitarSet is used within a six-fold cross-validation schema, with the dataset splits representing the tracks produced by each respective guitarist. We limit training to only four splits during each fold, holding out one split for validation and one split for testing.

\subsubsection{DadaGP}
\label{sec:dadagp}
{DadaGP} \cite{sarmento2021dadagp} is a large collection of symbolic tablature encoded using the proprietary \textit{GuitarPro} file format. The dataset features many popular songs from a variety of artists and spanning many musical styles, with a bias toward rock and metal music. The \textit{GuitarPro} file format can store the transcription of multiple musical voices as tracks in a single file. Tracks corresponding to guitars have note labels which are always associated with an S/F combination.

We process all tracks corresponding to guitars in standard tuning within the \textit{GuitarPro} files, ignoring duplicate\footnote{A duplicate is defined as having a preexisting file name or the ``copy'' tag in the file name. In cases where there are duplicates with alternate file extensions, we keep the one with the more recent \textit{GuitarPro} version.} files. Note that many files include more than one guitar track. Using the Python package \textit{PyGuitarPro} \cite{Abakumov2020pyguitarpro}, we carry out a series of steps\footnote{Due to the many complexities of \textit{GuitarPro}, we refer interested readers to the code for more information about these steps. It is also worth mentioning here that we treat slides and hammer-ons/offs as two separate notes, and that for bends we only retain the original note.} where we assign an onset and offset to each note in the track to obtain tablatures in \textit{JAMS} format \cite{humphrey2014jams}. Ultimately, we end up with 33967 pieces of symbolic tablature, most of which comes from full-length songs. We use these to estimate pairwise likelihoods.

\subsection{Metrics}
\label{sec:metrics}
We compute all of the same metrics\footnote{The original metrics were computed across all frames of all tracks in a track-agnostic manner, whereas we compute the metrics across all frames of each track and average results across all tracks. This is done to weight the influence of every track in the dataset equally.} as in \cite{wiggins2019guitar}. These include precision, recall, and f-measure for all frames of tablature and the equivalent string-agnostic multipitch, as well as the tablature disambiguation rate (TDR), which indicates how well the model maps pitches to their respective strings. We also report the average inhibition losses $L_{inh}$ and $L_{inh} {}^+$ using the standard ($b = 1$) and boosted ($b = 2^7$) inhibition weights, respectively. The inhibition losses serve as a proxy for how well the predictions of the model match the distribution of the symbolic tablature used to estimate the pairwise likelihoods. The lower the losses, the smaller the presence of inhibitory pairs in the predictions. Since the pairwise likelihoods were estimated using a large collection of real-world data, the inhibition losses also serve as a proxy for the feasibility of the predicted tablature. Note that these inhibition losses are calculated on the final predictions of a model. This is in contrast to the inhibition loss applied during training, which is computed on the activations after the sigmoid operation.

Additionally, we note that the inhibition weights tend to be very high for pairs on different strings corresponding to the same exact pitch $p \in \{1, ..., P\}$, where $P$ is the total number of unique pitches. A duplicate pitch error is a common mistake made by the network, since the acoustic profile of the same pitch on different strings is very similar. We report the average number of duplicate pitch errors per track, where the duplicate pitch error count $E_{d.p.}$ for a track with $N$ frames is expressed as
\begin{equation}
\label{eq:duplicate_pitch_error}
E_{d.p.} = \sum_{n = 1}^{N} \sum_{p = 1}^{P} \max(0, m^{(y)}_{p, n} - m^{(t)}_{p, n}),
\end{equation}
where $m^{(y)}_{p, n}$ $\in \{0, \dots, 5\}$ and $m^{(t)}_{p, n}$ $\in \{0, \dots, 5\}$ are the number of string-wise model predictions and ground-truth targets, respectively, at frame $n$ corresponding to pitch $p$. Since it is extremely rare that a guitarist duplicates the same pitch in practice, this error count is a more explicit way of telling if the inhibition objective is effective. We also report the average number of false alarm errors,
\begin{equation}
\label{eq:false_alarm_error}
E_{f.a.} = \sum_{n = 1}^{N} \sum_{s = 1}^{6} \sum_{f = 0}^{F} y_{s, f, n} \land \lnot \text{ } t_{s, f, n},
\end{equation}
indicating more generally all predictions made by the model which do not occur in the ground-truth. This allows us to compare the number of duplicate pitch errors to the total number of false alarm errors in each experiment. Note that here we do not consider incorrect silence predictions ($f = -1$) to be false alarm errors.

\subsection{Experiments}
\label{sec:experiments}
We conduct a series of experiments to observe how the proposed output layer formulation compares to the baseline 6D Softmax formulation. All of our experiments are trained and evaluated on GuitarSet following the six-fold cross-validation schema laid out in \cite{wiggins2019guitar}. There is one major difference in that we hold out one extra dataset split for validation. This means that per fold, only four splits are used for training, while the other two are used for validation and evaluation, respectively. Within each fold, the criterion for choosing the model checkpoint to evaluate on the evaluation split is the checkpoint with the maximum tablature f-measure on the validation split.

The purpose of the first few experiments is to verify our reproduction of TabCNN and our experimental setup, and to investigate the difference in performance when using a validation set. The models in these experiments are trained with a batch size of 30 for 10000 iterations, where 200 consecutive frame groups within each track in the training set are sampled per iteration. Experiment (1) features our reproduction of TabCNN, with the standard 6D Softmax output layer formulation. In order to try to match the original TabCNN results as closely as possible, we do not perform validation in this experiment, and simply evaluate the final models within each fold at 10000 iterations. Next, in experiment (2), we run the same experiment but with the validation methology outlined above. All of the remaining experiments utilize the same validation methodology.

The purpose of the remaining experiments is to compare the 6D Softmax and Logistic output layer formulations directly. As discussed in Sec. \ref{sec:baseline_model}, an LSTM is inserted before the output layer of each model in these experiments as a simple improvement, and to observe what happens to the proposed metrics when a simple language model is added. In order to create a balance between the sequence length and the batch size when training the models with the LSTM, we also modify the training hyperparameters such that we train with a batch size of 50 with a sequence length of 125 frame groups for 50000 iterations. We increase the amount of training iterations to account for the additional model complexity of the LSTM.

\begin{figure}[t]
\includegraphics[width=\columnwidth]{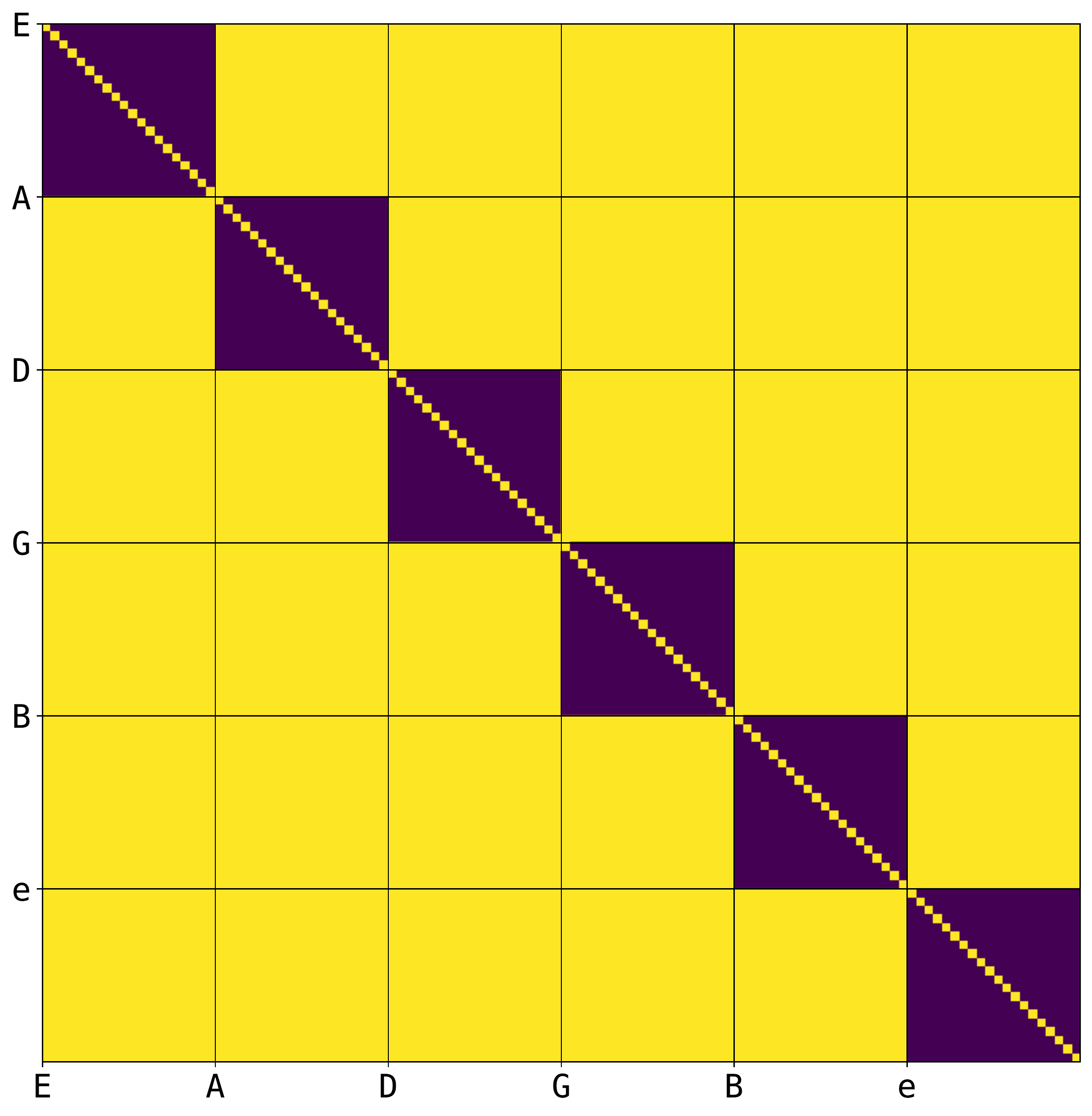}
\caption{Complement of the inhibition weights which enforce hard string constraints, described by Equation (\ref{eq:string_constraints}).}
\label{fig:strings_weight_complement}
\end{figure}

Experiment (3) is the same as Experiment (2), except for the insertion of the LSTM layer and the new training hyperparameters. Experiment (4) features the Logistic output layer formulation detailed in Sec. \ref{sec:output_layer_formulation}, but with an ineffective inhibition objective, \textit{i.e.} $\lambda = 0$. In the remaining experiments, $\lambda \neq 0$, and various different matrices of inhibition weights are employed. In Experiment (5), a set of weights which only cover the hard string constraints is used for the inhibition loss. The weights can be expressed as the following:
\begin{equation}
\label{eq:string_constraints}
w(c_i, c_j) = \left\{
    \begin{array}{lc}
        1 & \text{if } |i - j| < (F + 2) \text{ and } i \neq j \\
        0 & \text{otherwise}
    \end{array}
    \right. .
\end{equation}
This matrix of weights inhibits all of the pairs which are intrinsically impossible with the 6D Softmax formulation. A visualization of the complement to these weights is provided in Fig. \ref{fig:strings_weight_complement}.  All other experiments employ matrices estimated using DadaGP \cite{sarmento2021dadagp} with the procedure detailed in Sec. \ref{sec:estimating_pairwise_likelihood}. The standard inhibition weights ($DadaGP$) are used in Experiment (6). Experiment (7) and (8) use inhibition weights boosted with $b = 2^7$ ($DadaGP^+$). The estimated pairwise likelihoods corresponding to each set of weights are presented in Fig. \ref{fig:pairwise_likelihoods_dadagp}. Experiments (4-7) use an inhibition loss multiplier of $\lambda = 1$, whereas Experiment (8) uses $\lambda = 10$. These experiments make up a sort of ablation study w.r.t. the design choices of the inhibition objective within the Logistic formulation.

\section{Results \& Discussion}
\label{sec:results_and_discussion}
\begin{table*}[ht]
\begin{center}
  \begin{tabular}{|l||c|c|c||c|c|c||c|}
    \hline
    Tablature Layer & $\mathit{p_{tab}}$ & $\mathit{r_{tab}}$ & $\mathit{f_{tab}}$ & $\mathit{p_{pitch}}$ & $\mathit{r_{pitch}}$ & $\mathit{f_{pitch}}$ & $\mathit{TDR}$ \\
    \hline
    \hline
    \textit{(1) Reproduction} & $0.809$ & $0.692$ & $0.742$ & $0.910$ & $0.762$ & $0.825$ & $0.903$ \\
    \hline
    \textit{(2) Reproduction w/ Val.} & $0.775$ & $0.696$ & $0.730$ & $\textbf{0.895}$ & $0.781$ & $0.830$ & $0.886$ \\
    \hline
    \hline
    \textit{(3) Reproduction w/ Val./Rec.} & $0.783$ & $0.757$ & $0.768$ & $0.879$ & $0.835$ & $0.854$ & $0.905$ \\
    \hline
    \textit{(4) Logistic (No Inhibition)} & $0.782$ & $0.757$ & $0.767$ & $0.878$ & $0.836$ & $0.854$ & $0.902$ \\
    \hline
    \textit{(5) Logistic w/ String Constraints} & $\textbf{0.789}$ & $\textbf{0.761}$ & $\textbf{0.773}$ & $0.881$ & $\textbf{0.836}$ & $\textbf{0.856}$ & $\textbf{0.907}$ \\
    \hline
    \textit{(6) Logistic w/ DadaGP} & $0.787$ & $0.743$ & $0.763$ & $0.880$ & $0.821$ & $0.847$ & $0.902$ \\
    \hline
    \textit{(7) Logistic w/ DadaGP$^+$} & $0.782$ & $0.754$ & $0.766$ & $0.876$ & $0.833$ & $0.852$ & $0.902$ \\
    \hline
    \textit{(8) Logistic w/ DadaGP$^+$\small$(\lambda=10)$} & $0.781$ & $0.755$ & $0.766$ & $0.867$ & $0.829$ & $0.845$ & $0.907$ \\
    \hline
  \end{tabular}
\end{center}
\caption{Average six-fold cross validation results on GuitarSet \cite{quingyang2019guitarset} for transcription metrics. Bold values indicate the highest observed result for each metric across experiments which follow the validation methodology outlined in Sec. \ref{sec:experiments} (\textit{i.e.,} Experiments (2-8)). The break separates experiments without recurrence and experiments which used an LSTM.}
\label{tab:results_transcription_metrics}
\end{table*}
\begin{table*}[ht]
\begin{center}
  \begin{tabular}{|l||c|c|c|c|}
    \hline
    Tablature Layer & $L_{inh}$ & $L_{inh} {}^+$ & $E_{d.p.}$ & $E_{f.a.}$ \\
    \hline
    \hline
    \textit{(1) Reproduction} & $8.87$ & $0.132$ & $21.4$ & $359.8$ \\
    \hline
    \textit{(2) Reproduction w/ Val.} & $\textbf{9.01}$ & $0.152$ & $34.2$ & $\textbf{442.5}$ \\
    \hline
    \hline
    \textit{(3) Reproduction w/ Val./Rec.} & $9.27$ & $0.158$ & $24.3$ & $489.6$ \\
    \hline
    \textit{(4) Logistic (No Inhibition)} & $9.27$ & $0.154$ & $20.0$ & $503.3$ \\
    \hline
    \textit{(5) Logistic w/ String Constraints} & $9.25$ & $0.155$ & $19.5$ & $485.8$ \\
    \hline
    \textit{(6) Logistic w/ DadaGP} & $9.19$ & $0.147$ & $12.0$ & $481.8$ \\
    \hline
    \textit{(7) Logistic w/ DadaGP$^+$} & $9.25$ & $0.143$ & $13.8$ & $496.6$ \\
    \hline
    \textit{(8) Logistic w/ DadaGP$^+$\small$(\lambda=10)$} & $9.26$ & $\textbf{0.132}$ & $\textbf{10.6}$ & $504.6$ \\
    \hline
  \end{tabular}
\end{center}
\caption{Average six-fold cross validation results on GuitarSet \cite{quingyang2019guitarset} for distribution and error metrics. Bold values indicate the lowest observed result for each metric across experiments which follow the validation methodology outlined in Sec. \ref{sec:experiments} (\textit{i.e.,} Experiments (2-8)). The break separates experiments without recurrence and experiments which used an LSTM.}
\label{tab:results_distribution_metrics}
\end{table*}

\begin{figure}[ht]
\includegraphics[width=\columnwidth]{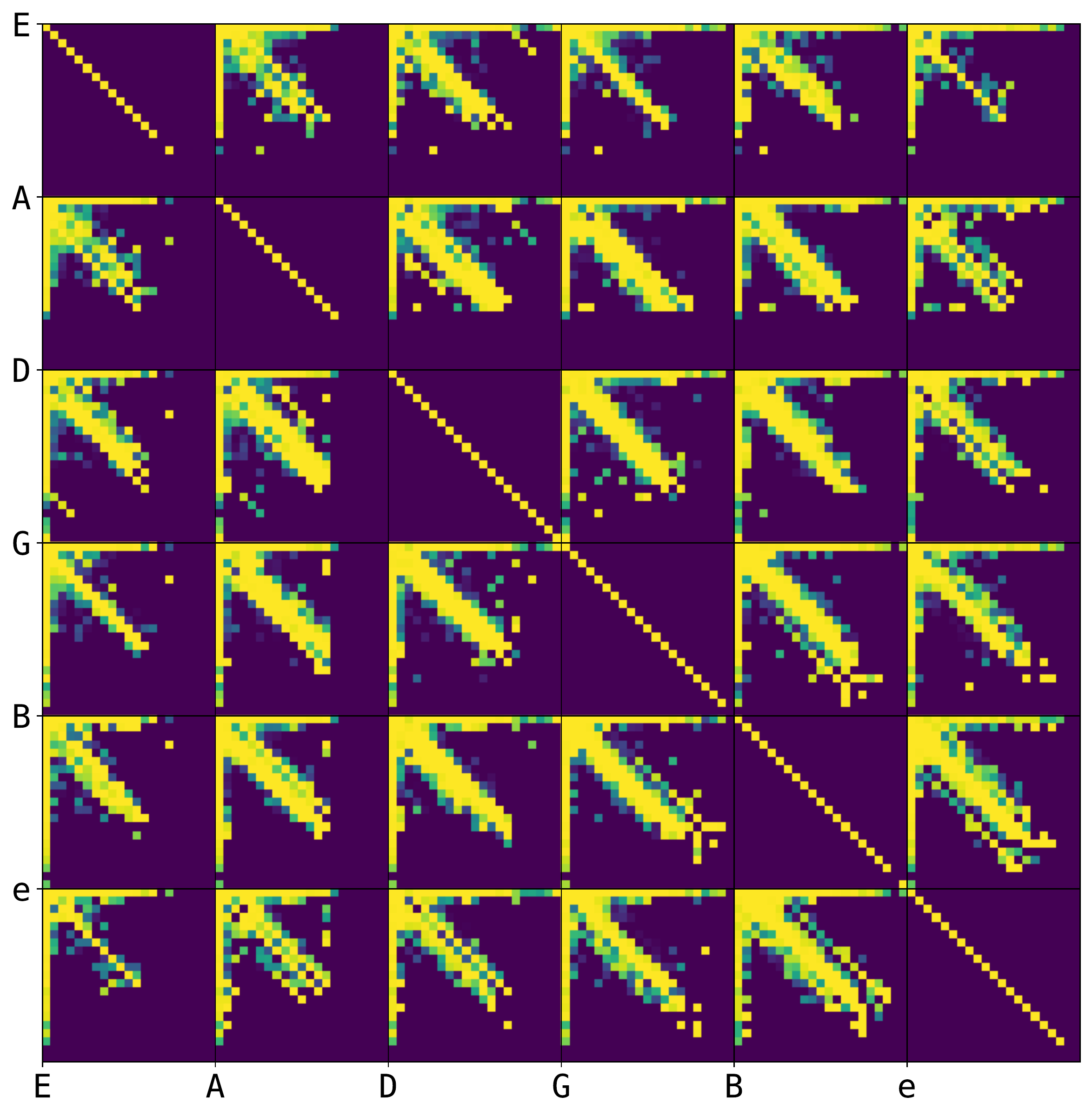}
\caption{Estimated pairwise likelihoods for all S/F combinations computed using GuitarSet \cite{quingyang2019guitarset} with $b = 2^7$.}
\label{fig:pairwise_likelihoods_guitarset}
\end{figure}

\subsection{Transcription}
The transcription results for the experiments outlined in Sec. \ref{sec:experiments} are presented in \tabref{tab:results_transcription_metrics}. The results of Experiment (1), our reproduction of TabCNN \cite{wiggins2019guitar}, are almost identical to what was originally reported. Regarding Experiment (2), validation lowers transcription performance slightly, as expected, but to a surprisingly small degree. This validates our decision to trade an extra dataset split for a more justifiable selection criterion. In Experiment (3), insertion of the LSTM significantly improves tablature transcription performance, mainly by increasing recall.

Without the inhibition objective, the Logistic formulation in Experiment (4) yields essentially the same performance as the 6D Softmax formulation in Experiment (3). When using the inhibition objective with the string constraints in Experiment (5), the overall tablature transcription performance improves slightly. Since the weights only discourage activations on the same string from co-occurring, similar to the 6D Softmax formulation, this is an interesting result. With the $DadaGP$ weights in Experiment (6), the inhibition objective lowers performance slightly by decreasing tablature and multipitch recall. This is not surprising, since the weights are very strict and inhibit almost everything besides perfect $4^{th}$/$5^{th}$ intervals. With the boosted weights $DadaGP^+$ in Experiments (7-8), recall improves and precision drops slightly for tablature and multipitch. Finally, the stronger inhibition objective $(\lambda=10)$ in Experiment (8) lowers overall performance for multipitch estimation, while maintaining roughly the same tablature transcription performance w.r.t. Experiment (7).

The lack of an increase in tablature performance when using the inhibition objective can most likely be attributed to the small size of GuitarSet \cite{quingyang2019guitarset} and the presence of some noisy labels. Using the procedure detailed in Sec. \ref{sec:estimating_pairwise_likelihood}, pairwise likelihoods were estimated from GuitarSet \cite{quingyang2019guitarset} for analysis. These are illustrated in Fig. \ref{fig:pairwise_likelihoods_guitarset}. Clearly, there are co-occurrences for pairs which are highly inhibited by the $DadaGP$ and $DadaGP^+$ weights. Upon inspection of the tracks which produced these artifacts, we found some instances of likely duplicate pitch and octave errors in the annotations. It could be that the inhibition variants actually avoided making these types of predictions, but received a lower score due to some annotation errors. Furthermore, the rock and metal bias in DadaGP \cite{sarmento2021dadagp} may have skewed the distribution of model predictions in the relevant models away from the distribution of GuitarSet \cite{quingyang2019guitarset}, which contains genres such as Jazz and Bossa Nova.

\subsection{Loss \& Errors}
The distribution and error measurements for the experiments outlined in Sec. \ref{sec:experiments} are presented in \tabref{tab:results_distribution_metrics}. Some key observations can be made about these results. First, the standard inhibition loss $L_{inh}$ remains relatively consistent across experiments. As noted earlier, the $DadaGP$ weights are very strict and inhibit most pairs, so the inhibition objective can conflict with the transcription objective. For this reason, it makes sense that among Experiments (3-8), Experiment (6) had the most influence on this metric, given that it used $L_{inh}$ for training. We also observe a significant decrease in $L_{inh} {}^+$ for Experiments (7-8), which train with the boosted inhibition loss. In contrast to $L_{inh}$, even when $L_{inh} {}^+$ continues to go down, the transcription performance improves for Experiments (6-8). This agrees with our hypothesis that the boosted inhibition weights are more suitable for use with the transcription objective.

We also notice that, despite a significant increase in Experiment (2) and subsequently Experiment (3), the average number of false alarm errors $E_{f.a.}$ remains relatively consistent across Experiments (3-8). It does seem to improve slightly with inhibition using string constraints and the $DadaGP$ weights. The average number of duplicate pitch errors $E_{d.p.}$ improves significantly in Experiments (6-8). As discussed before, there are cases of duplicate pitches in the ground-truth of GuitarSet. The reason why the duplicate pitch error count is not lower may be because the models are trained to produce duplicate pitch predictions in some scenarios. Overall, we argue that the lower $E_{d.p.}$ and $L_{inh} {}^+$ suggests that models trained with $DadaGP$ and $DadaGP^+$ produce tablature which is more feasible to play and more consistent with DadaGP \cite{sarmento2021dadagp}.

\subsection{Future Work}
Although the inhibition objective was presented here in the context of improving guitar tablature transcription, it has several other potential uses. Inhibition shapes the distribution of model predictions by inhibiting unlikely S/F pairs. This means that one can estimate the pairwise likelihoods using a curated distribution, \textit{e.g. }a collection of tablature corresponding to a specific musician or genre. Similarly, collections based on playing difficulty could be curated to influence the model to produce tablature more suitable for specific users with varying proficiency.

Another usage could be within the context of tablature arrangement, where the inhibition objective could be applied to train a model to allocate a set of preexisting pitches to strings, such that the resulting fingerings are playable. This may even be useful in a two-stage approach to tablature transcription, where a generic multipitch estimation model feeds predictions into the arrangement system.

We also suggest several directions for improving the inhibition objective. First, one could explore various types of data augmentation for symbolic tablature. One example we refer to as capo augmentation, where a constant fret offset is added to all notes within a track. This could prevent issues related to a dataset's lack of fretboard coverage. Another interesting direction would be the inclusion of a temporal inhibition objective at the note-level. This could prevent a model from generating predictions which shift around the fretboard too much. Lastly, it would be interesting to investigate higher-order S/F relationships (\textit{e.g.}, 3 notes or more), since the current method only takes pairwise relationships into account.

\section{Conclusion}
\label{sec:conclusion}
We propose a new output layer formulation for guitar tablature transcription which takes advantage of large collections of symbolic tablature data. The pairwise likelihood of concurrent activation for all possible notes on the guitar is estimated using a recently published dataset. The complement of the pairwise likelihood is used as a weight for an accompanying inhibition objective during training. We compare the new formulation against the output layer formulation of a baseline tablature transcription model. The inhibition objective is shown to be effective in shaping the distribution of the output predictions and lowering the number of duplicate pitch errors. We also discuss alternative uses and future directions for the inhibition objective.

\begin{acknowledgments}
This work has been partially funded by the National Science Foundation grants IIS-1846184 and DGE-1922591.
\end{acknowledgments} 

\bibliography{paper}

\end{document}